\documentclass[12pt]{iopart}

\usepackage{listings}
\usepackage{graphicx}
\usepackage{iopams}
\usepackage{cite}
\usepackage{hyperref}

\newcommand{\fstat}{$\mathcal{F}$-statistic}

\newcommand{\dd}{\mathrm{d}}
\newcommand{\abs}[1]{|#1|}

\begin{document}

\title[GPU $\mathcal{F}$-statistic]{Graphics processing unit implementation of the $\mathcal{F}$-statistic for continuous gravitational wave searches}

\author{Liam Dunn$^{1,2}$, Patrick Clearwater$^{3,1,2}$, Andrew Melatos$^{1,2}$, and Karl Wette$^{4,2}$}

\address{$^1$ School of Physics, University of Melbourne, Parkville VIC 3010, Australia}
\address{$^2$ ARC Centre of Excellence for Gravitational Wave Discovery (OzGrav), Hawthorn VIC 3122, Australia}
\address{$^3$ Gravitational Wave Data Centre, Swinburne University of Technology, Hawthorn VIC 3122, Australia}
\address{$^4$ Centre for Gravitational Astrophysics, Australian National University, Canberra ACT 2601, Australia}

\ead{liamd@student.unimelb.edu.au}

\begin{abstract}
The $\mathcal{F}$-statistic is a detection statistic used widely in searches for continuous gravitational waves with terrestrial, long-baseline interferometers.
A new implementation of the $\mathcal{F}$-statistic is presented which accelerates the existing ``resampling'' algorithm using graphics processing units (GPUs).
The new implementation runs between 10 and 100 times faster than the existing implementation on central processing units without sacrificing numerical accuracy.
The utility of the GPU implementation is demonstrated on a pilot narrowband search for four newly discovered millisecond pulsars in the globular cluster Omega Centauri using data from the second Laser Interferometer Gravitational-Wave Observatory observing run.
The computational cost is $17.2$ GPU-hours using the new implementation, compared to 1092 core-hours with the existing implementation.
\end{abstract}

\submitto{\CQG}

\maketitle

\section{Introduction} \label{sec:intro}
Gravitational waves emitted by isolated and accreting neutron stars \cite{Riles2017} are a key target for terrestrial, long-baseline interferometers, such as the Laser Interferometer Gravitational-wave Observatory (LIGO) \cite{2015lscadvligo}, Virgo
\cite{2015acernese}, and the Kamioka Gravitational Wave Detector (KAGRA) \cite{KAGRACollaboration2019}. Individual neutron stars are continuous wave sources; they emit persistent harmonic signals as they rotate.
Continuous wave searches are typically limited in sensitivity and parameter range by available computational resources. This is particularly true when the astrophysical parameters (e.g. the neutron star spin frequency and/or binary orbital elements) are poorly constrained by electromagnetic observations.
Continuous wave sources are relatively weak compared to the compact binary coalescences which have been detected already.
These coalescences have typical peak gravitational-wave strains of order $10^{-22}$ \cite{Abbott2016, Abbott2017b}, while the current most stringent 95\% confidence level upper limits on the gravitational-wave strain from continuous wave sources are of order $10^{-25}$ for all-sky searches with wide parameter spaces \cite{DergachevPapa2021, AbbottAbbott2021}, and of order $10^{-26}$ for targeted searches for emission from known pulsars \cite{AshokBeheshtipour2021, AbbottAbbott2020}.
Sensitivity to these weak continuous wave sources is enhanced by integration over long time spans (hundreds of days), posing a further computational challenge.

A promising approach for dealing with computationally-heavy problems is to reformulate them to take advantage of graphics processing units (GPUs). GPUs can execute thousands of tasks in parallel, in contrast to the dozens of tasks which can be performed by a modern, multi-core central processing unit (CPU).
They are particularly well-suited to ``embarrassingly parallel'' problems, which can be broken down into simple sub-steps, where each sub-step performs the same instructions on different data, and few sub-steps depend on the results of other sub-steps.
Frameworks such as \texttt{CUDA} \cite{NvidiaCorporation} and \texttt{OpenCL} \cite{Stone2010}, which enable general-purpose computing on GPUs, have unleashed the power of GPUs on a range of applications in industry and science.
In the field of gravitational wave astronomy in particular, GPUs have been brought to bear on problems such as the low-latency detection of compact binary coalescences \cite{Liu2012, Guo2018}, Bayesian inference applied to compact binary coalescences \cite{Talbot2019, Wysocki2019}, searches for long-lived transient signals \cite{Keitel2018}, and a search for continuous gravitational waves from Scorpius X-1 \cite{TheLIGOScientificCollaboration2019}.
Recently, a new GPU-based implementation of the FrequencyHough algorithm \cite{AstoneColla2014} for all-sky continuous wave searches has been reported \cite{LaRosaAstone2021}, achieving a speed-up of 1--2 orders of magnitude over the existing CPU-based implementation. 
The application of deep learning techniques to the analysis of both compact binary coalescences \cite{George2018, Gabbard2018} and continuous gravitational waves \cite{Dreissigacker2019} has also benefited greatly from the computing power provided by GPUs.

One of the principal tools employed in continuous-wave searches is the $\mathcal{F}$-statistic, introduced by \cite{Schutz1998}.
This is a maximum-likelihood technique requiring numerical maximisation over certain parameters which control the evolution of the phase of the signal at the detector.
It has been applied to searches for electromagnetically discovered targets at known sky locations \cite{Abbott2019Known} and all-sky surveys for unknown targets \cite{Abbott2019}.
The LIGO Algorithms Library Suite (\texttt{LALSuite}) \cite{lalsoft} currently contains two implementations of the \fstat{} \cite{Prix2011}: the demodulation algorithm (\texttt{Demod}) \cite{WilliamsSchutz2000}, which is older and typically runs slower; and the resampling algorithm (\texttt{Resamp}) \cite{Patel2009}, which typically runs faster.
Both implementations operate on CPUs.

In this paper we present a new implementation of the resampling algorithm, which offloads most of the computation to GPUs, achieving substantial speed-up over the CPU implementation.
The paper is structured as follows.
In section \ref{sec:fstat} we review the mathematical definition of the $\mathcal{F}$-statistic, so that it is clear what data it accepts as input and what arithmetic and logical operations it entails.
In section \ref{sec:impl} we outline the algorithmic steps involved in calculating the $\mathcal{F}$-statistic and explain how to implement these steps within the \texttt{CUDA} framework for GPU programming.
In section \ref{sec:perf} we evaluate the runtime (section \ref{subsec:runtime}) and accuracy (section \ref{subsec:acc}) of the GPU implementation, and compare against the CPU implementations available in \texttt{LALSuite}.
In section \ref{sec:pilot} we describe a pilot search using the GPU implementation to target four newly discovered millisecond pulsars in the globular cluster Omega Centauri \cite{daiDiscoveryMillisecondPulsars2020}, using data from the second LIGO-Virgo observing run (O2).
Neutron stars in Omega Centauri have not been the subject of any targeted or directed searches for continuous gravitational waves previously.
Nevertheless we emphasise that the search described here is only a proof of principle demonstrating the utility of the GPU implementation of the \fstat{}, not a full-scale astrophysical search.
Finally in section \ref{sec:discussion} we discuss future prospects for additional optimisation and the implications for future \fstat{} searches with LIGO data.

\section{$\mathcal{F}$-statistic}\label{sec:fstat}
The \fstat{} is a detection statistic which is constructed from the likelihood ratio
\begin{equation}
    \mathcal{L}(x; \mathcal{A}, \lambda) \equiv \frac{\Pr[x \mid \mathcal{H}_S(\mathcal{A},\lambda)]}{\Pr[x \mid \mathcal{H}_N]}.
\end{equation}
Conditional probabilities are denoted $\Pr[\cdot\!\mid\!\cdot]$, and following the notation of \cite{prixTargetedSearchContinuous2009}, the observed data are denoted by $x$, $\mathcal{A}$ is the set of ``amplitude parameters'' determining the amplitude of the signal at the detector, and $\lambda$ is the set of ``phase parameters'' which determine the evolution of the phase of the signal at the detector.
The two hypotheses are $\mathcal{H}_N$, the hypothesis that the data consist solely of Gaussian noise with known power spectral density, and $\mathcal{H}_S(\mathcal{A},\lambda)$, the hypothesis that the data consist of Gaussian noise with known power spectral density, plus a continuous wave signal with parameters $\mathcal{A}, \lambda$.
The  likelihood ratio $\mathcal{L}(x; \mathcal{A}, \lambda)$ can be analytically maximised over $\mathcal{A}$, leaving the maximum-likelihood function
\begin{equation}
    \mathcal{L}_{\mathrm{ML}}(x; \lambda) = \max_{\mathcal{A}} \mathcal{L}(x; \mathcal{A}, \lambda) = e^{\mathcal{F}(x, \lambda)}. \label{eqn:fstat_defn}
\end{equation}
The final equality in (\ref{eqn:fstat_defn}) serves as the definition of $\mathcal{F}(x, \lambda)$ (hereafter written as $\mathcal{F}$ for brevity).

In practice $\mathcal{F}$ is calculated via two Fourier transforms of the data, weighted by the two antenna beam pattern functions of the detector.
The phase of a continuous wave signal from a rotating neutron star can be written as \cite{Schutz1998}
\begin{equation}
    \Phi(t) = 2\pi f_0[t + t_m(t; \alpha, \delta)] + \Phi_s\!\left[t; f_0^{(k)}, \alpha, \delta\right],
\end{equation}
where $t_m$ gives the time offset due to the motion of the detector, and $\Phi_s$ gives the phase due to the rotational evolution of the source.
Here $\alpha$ and $\delta$ are the right ascension and declination of the source, $f_0$ is the gravitational-wave frequency, $f_0^{(k)}$ is the $k$-th time derivative of the gravitational-wave frequency, and $\Phi_s$ is expressed as a Taylor series in time, whose coefficients are given by $f_0^{(k)}$ ($k \geq 1$).
Following \cite{Schutz1998} throughout, we define the two integrals
\begin{eqnarray}
    F_{a}(f_0) &= \int_{-T_\mathrm{obs}/2}^{T_\mathrm{obs}/2}\dd t\,x(t) a(t) \exp[-i\Phi_s(t)]\exp\{-i2\pi f_0[t + t_m(t)]\}, \label{eqn:Fa_DET} \\
    F_{b}(f_0) &= \int_{-T_\mathrm{obs}/2}^{T_\mathrm{obs}/2}\dd t\, x(t) b(t) \exp[-i\Phi_s(t)]\exp\{-i2\pi f_0[t + t_m(t)]\}, \label{eqn:Fb_DET}
\end{eqnarray}
where $T_\mathrm{obs}$ is the total observing time, $x(t)$ is the data time series in the detector frame, and $a(t)$ and $b(t)$ encode the beam-pattern functions, also in the detector frame.
Explicit formulas for $a(t)$ and $b(t)$ are given by equations (12) and (13) of \cite{Schutz1998} in terms of $\alpha$, $\delta$, and the detector's orientation.
Introducing a new barycentred time coordinate, $t_b(t) = t + t_m(t)$, and making the approximations $t_b(T_\mathrm{obs}/2) \approx T_\mathrm{obs}/2$ and $\dd t/\dd t_b \approx 1$, we rewrite (\ref{eqn:Fa_DET}) and (\ref{eqn:Fb_DET}) as
\begin{eqnarray}
    F_a(f_0) = \int_{-T_\mathrm{obs}/2}^{T_\mathrm{obs}/2} \dd t_b\, x[t(t_b)] a[t(t_b)] \exp\{-i\Phi_s[t(t_b)]\}\exp(-i2\pi f_0 t_b), \label{eqn:Fa} \\
    F_b(f_0) = \int_{-T_\mathrm{obs}/2}^{T_\mathrm{obs}/2} \dd t_b\, x[t(t_b)] b[t(t_b)] \exp\{-i\Phi_s[t(t_b)]\}\exp(-i2\pi f_0 t_b).  \label{eqn:Fb}
\end{eqnarray}
The detection statistic $\mathcal{F}$ is defined as
\begin{equation}
    \mathcal{F} = \frac{4}{S_h(f_0)T_\mathrm{obs}}\frac{B\abs{F_a}^2 + A\abs{F_b}^2 - 2C\Re(F_aF_b^*)}{D}, \label{eqn:Fstat}
\end{equation}
where $A, B, C$, and $D = AB - C^2$ are integrals of products of $a(t)$ and $b(t)$ over $-T_\mathrm{obs}/2 \leq t \leq T_\mathrm{obs}/2$ (i.e. constants depending on $\alpha$ and $\delta$), and $S_h(f_0)$ is the one-sided power spectral density of the detector noise at the frequency $f_0$.

Equations (\ref{eqn:Fa}) and (\ref{eqn:Fb}) are standard Fourier transforms of the data $x[t(t_b)]$ multiplied by the slowly varying  functions $a[t(t_b)]$ and $b[t(t_b)]$, with $\abs{\dot{a}} \ll f_0\abs{a}$ and $\vert\dot{b}\vert \ll f_0\abs{b}$. Hence they can be computed efficiently using standard CPU and GPU libraries, e.g. \texttt{FFTW} and \texttt{cuFFT} \cite{press2007numerical}.

\section{GPU Implementation}\label{sec:impl}
We develop a \texttt{CUDA} implementation of the \fstat{} based on the \texttt{Resamp} algorithm, suggested by \cite{Schutz1998} and elaborated and implemented by \cite{Patel2009}, whose notation we follow.
A mature CPU-based implementation of \texttt{Resamp} already exists in \texttt{LALSuite}. The \texttt{CUDA} implementation is a fairly direct ``translation'' of the CPU code.
It hooks in smoothly to the rest of the \texttt{LALSuite} infrastructure. 
In particular, it is easily usable with the \texttt{ComputeFStatistic\_v2} program, which wraps the various \fstat{} routines available in \texttt{LALPulsar} (the portion of \texttt{LALSuite} containing routines for continuous GW data analysis), and provides logic for parameter searches over $f_0, f_0^{(k)}, \alpha$, and $\delta$ for example.

In practice, the starting point for analysis is not the raw time series $x(t)$ but rather a sequence of short Fourier transforms (SFTs).
These are discrete Fourier transforms of the detector output over time segments which are short enough (typically $T_{\mathrm{SFT}} \sim 1800\,\mathrm{s}$) that amplitude and phase modulations due to the Earth's motion and the source's frequency evolution are small.

The \texttt{Resamp} method consists of the following steps for a single detector; the generalization to multiple detectors is straightforward.
\begin{enumerate}
    \item Combine the SFTs into a single Fourier transform.
    \item Complex heterodyne at a frequency $f_h$, downsample, and low-pass filter the data in the frequency domain. Perform an inverse Fast Fourier Transform (FFT) to obtain the time series $x_h(t^k)$, where $k$ goes from 1 to the total number of samples, and $t^k$ is regularly spaced in the detector frame.
    \item Correct $x_h(t^k)$ to account for the fact that the heterodyning operation occurs in the detector frame rather than the Solar System barycentre frame.
    The result, $z(t^k) = x_h(t^k)e^{i 2 \pi f_h t_m}$, is heterodyned in the Solar System barycentre frame but is sampled irregularly in that frame.
    \item Choose $t^{k}_b$, a series of regularly spaced epochs in the Solar System barycentre frame. Calculate $T^{k}(t^{k}_b)$, the epochs in the detector frame corresponding to regular sampling in the Solar System barycentre frame.
    \item Calculate $z{\left[T^{k}(t^{k}_b)\right]}$ by interpolating $z(t^k)$.
    \item Calculate $F_a(f - f_h)$ and $F_b(f - f_h)$, by computing the FFTs of $a{\left[T^k(t^k_b)\right]}z{\left[T^k(t^k_b)\right]}\exp\left\{-i\Phi_s\left[T^k(t^k_b)\right]\right\}$ and $b{\left[T^k(t^k_b)\right]}z{\left[T^k(t^k_b)\right]}\exp\left\{-i\Phi_s\left[T^k(t^k_b)\right]\right\}$.
    \item Calculate $\mathcal{F}(f - f_h)$ from (\ref{eqn:Fstat}).
\end{enumerate}
Note that a single pass of the procedure (i)--(vii) can be used to compute $\mathcal{F}$ over a user-defined frequency band of width $\Delta f$, centred on $f_h$. 

The process of converting a sequence of SFTs into the time series $x_h(t^k)$ is complex (see Section III B of \cite{Patel2009} for details) and does not present a significant computational bottleneck, so we do not reimplement it. 
It is instead performed using the existing CPU-bound code in \texttt{LALSuite}. 
Steps (iii), (iv), (v), and (vii) can be expressed neatly as a simple operation or short sequence of operations executed independently for a large number of data points, where these data points may be elements of the detector data time series (either in the detector frame or the solar system barycentre frame), or frequency bins in the $\mathcal{F}$ results.
It is straightforward to write one or two \texttt{CUDA} kernels per step, which carry out the computations in parallel.
GPUs are extremely well-suited to performing the FFTs in step (vi).
An optimised FFT library for \texttt{CUDA} exists in \texttt{cuFFT}\footnote{\url{https://developer.nvidia.com/cufft}}, which we use here.

Steps (iii) through (vii) must be repeated for every search template [i.e. every combination of $\alpha, \delta$, and $f_0^{(k)}$ with $k \geq 1$], so accelerating them with GPUs yields computational savings which scale with the number of templates. 
In contrast, computing $x_h(t^k)$ using GPUs would deliver a speed-up which is constant with the number of templates.
For now we are content to amortise the cost of calculating $x_h(t^k)$ over the typically large number of templates, so that the proportion of computation performed on the CPU decreases as the number of templates increases.
The CPU time taken to perform steps (i) and (ii) is typically on the order of seconds to a few minutes depending on the nature of the data to be loaded, while the GPU time spent performing steps (iii) to (vii) may be on the order of hours to days, when the number of templates is large.
We note also that the transfer of data from CPU memory to GPU memory occurs only between steps (ii) and (iii) and is not repeated for every template, so this cost (which might otherwise be substantial) is also amortised.
The fact that all of the template-dependent computation occurs on the GPU means that searches over large numbers of templates can be carried out without introducing any inefficiencies due to communication between the CPU and GPU.
Thus searches over sky position $\alpha, \delta$ and spin-down parameters $f_0^{(k)}$ incur no extra performance penalty beyond the run time increasing in proportion to the number of templates.
We are able to use all of the existing infrastructure in \texttt{LALSuite} which manages these parameter space searches.

\begin{figure}
    \centering
    \includegraphics[width=\textwidth]{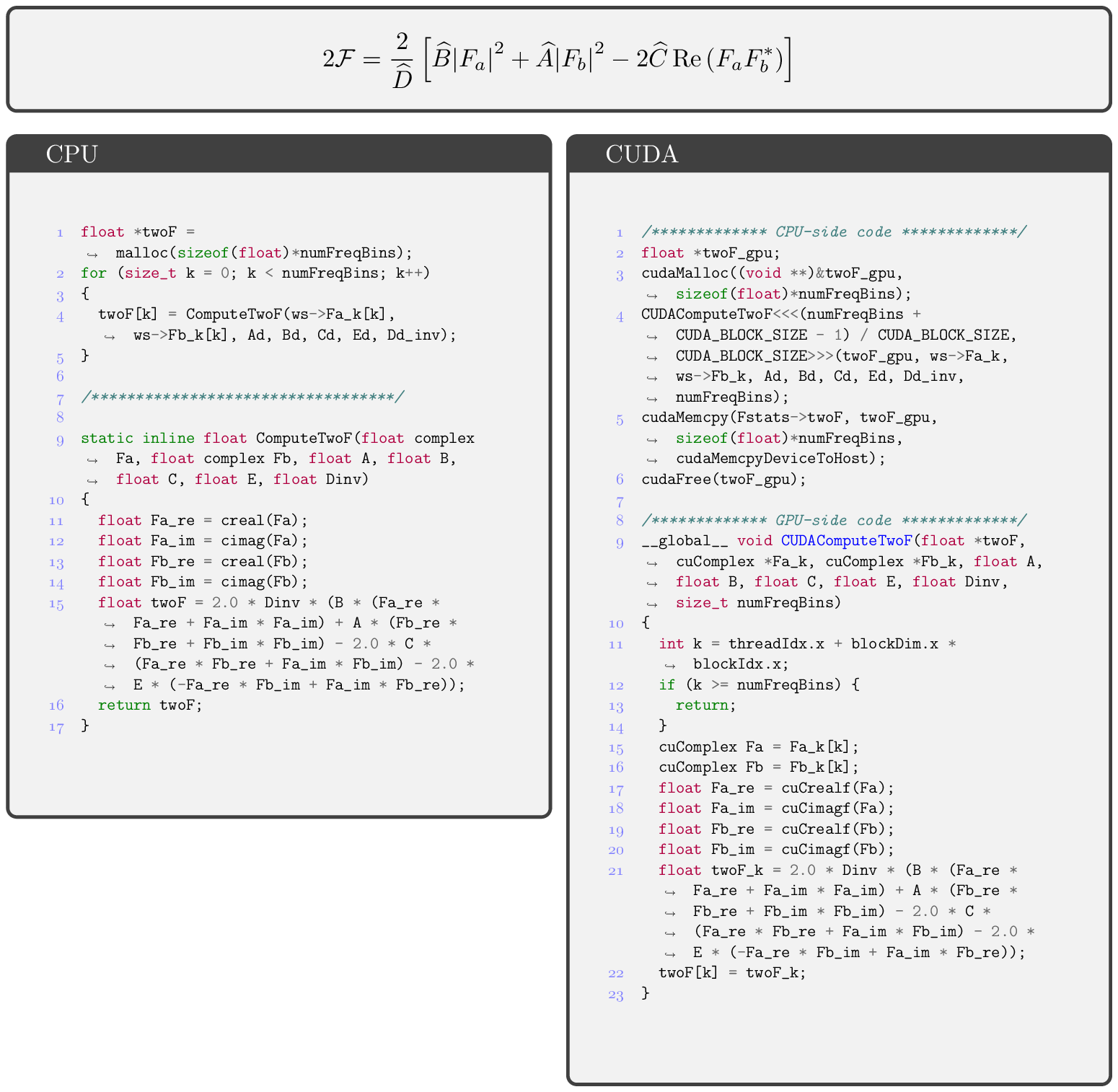}
    \caption{Side-by-side comparison of CPU code (left) and \texttt{CUDA} code (right) used to calculate $2\mathcal{F}$ according to equation (134) from \cite{Prix2011} (top), which is a rewriting of (\ref{eqn:Fstat}).
    The quantities $\widehat{A}$, $\widehat{B}$, $\widehat{C}$, $\widehat{D}^{-1}$ are normalised versions of the $A$, $B$, $C$, and $D^{-1}$ that appear in (\ref{eqn:Fstat}). Their definitions can be found in equation (133) of \cite{Prix2011}.
    The variable names in the code snippets are mostly self-explanatory, e.g. \texttt{Fa\_re} is the real part of $F_a(f_0)$ defined in equation \ref{eqn:Fa_DET}, and \texttt{Dinv} is $\widehat{D}^{-1}$. 
    The parameter \texttt{E} is zero in the limit where the arm length of the detector is much smaller than the gravitational-wave wavelength, which is always the case in the application considered here. Note that the function \texttt{ComputeTwoF} in the CPU code takes the name \texttt{compute\_fstat\_from\_fa\_fb}  in the LIGO Algorithm Library.}
    \label{fig:listing_combined}
\end{figure}

To give some idea of the similarities and differences between the CPU- and GPU-based implementations, figure \ref{fig:listing_combined} shows a side-by-side comparison of code snippets from the existing CPU code and the new \texttt{CUDA} code that perform step (vii) of the above procedure.

\section{Performance}\label{sec:perf}
To evaluate the performance of the \texttt{CUDA} implementation, we use the OzSTAR computing cluster\footnote{\url{https://supercomputing.swin.edu.au/ozstar/}}, with compute nodes equipped with Intel Gold 6140 CPUs running at 2.30 GHz (${\sim} 10^{11}$ floating point operations (FLOPs) per second per core), and Nvidia P100 GPUs (${\sim} 10^{13}$ FLOPs per second).
The benchmarking data were generated using the \texttt{Makefakedata\_v4} program in \texttt{LALSuite}.

We measure the performance of both the CPU and GPU implementations as a function of two key parameters: the total coherent observing time $T_\mathrm{obs}$, and the width of the frequency band $\Delta f$.
These two parameters control the number of frequency bins in the computation, which is the main factor determining the performance of the two implementations.
Varying other aspects of the computation, such as the number of terms in the Taylor series for the signal phase model, has no discernible effect on the performance of either implementation.
We vary only one parameter at a time, keeping $\Delta f$ fixed at $0.606\,\mathrm{Hz}$ as $T_{\mathrm{obs}}$ is varied, and keeping $T_\mathrm{obs}$ fixed at $10\,\mathrm{d}$ as $\Delta f$ is varied.
These choices are illustrative and are typical of certain published, $\mathcal{F}$-statistic-based searches for accreting neutron stars, e.g. \cite{Suvorova2016, TheLIGOScientificCollaboration2017,Suvorova2017, TheLIGOScientificCollaboration2019}.
For each parameter we search over 100 sky locations $(\alpha, \delta)$ and average the timing results. 
All other parameters used in the synthetic data generation and $\mathcal{F}$-statistic computation are held fixed.
The sky location does not noticeably affect the speed of the computation, and provides a convenient way to average the timing results over many executions of the algorithm without repeatedly incurring the start-up costs associated with steps (i) and (ii) described in the previous section.

We caution that the range of possible $\Delta f$ and $T_\mathrm{obs}$ combinations is limited by the amount of memory available on the GPU.
The scaling of required memory with both $\Delta f$ and $T_\mathrm{obs}$ is approximately linear (with deviations due to padding to ensure that the FFTs are carried out on data with lengths which are powers of two).
As an indicative example, the memory required for $\Delta f =0.606\,\mathrm{Hz}$, $T_{\mathrm{obs}} = 10\,\mathrm{d}$ is approximately $100\,\mathrm{MB}$, a modest amount compared to the several GB which are available on typical GPUs.
However, larger frequency bands and longer coherence times may present issues: increasing both $\Delta f$ and $T_\mathrm{obs}$ by a factor of 10 requires approximately $10\,\mathrm{GB}$ of GPU memory, approaching the limits of the Nvidia P100 GPUs used in testing which have $12\,\mathrm{GB}$ of memory.

\subsection{Runtime}\label{subsec:runtime}
We quote the performance results in terms of the timing coefficient $\tau_\mathcal{F}^\mathrm{eff}$  introduced by \cite{Prix2017}, defined as the time taken to calculate $\mathcal{F}$ for a single frequency bin in the single-detector case, starting from the downsampled, complex heterodyned, and filtered time series.
We measure $\tau_\mathcal{F}^\mathrm{eff}$ for the new \texttt{CUDA} implementation and the existing  CPU-bound implementation.
\begin{figure}
    \centering
    \includegraphics[width=\columnwidth]{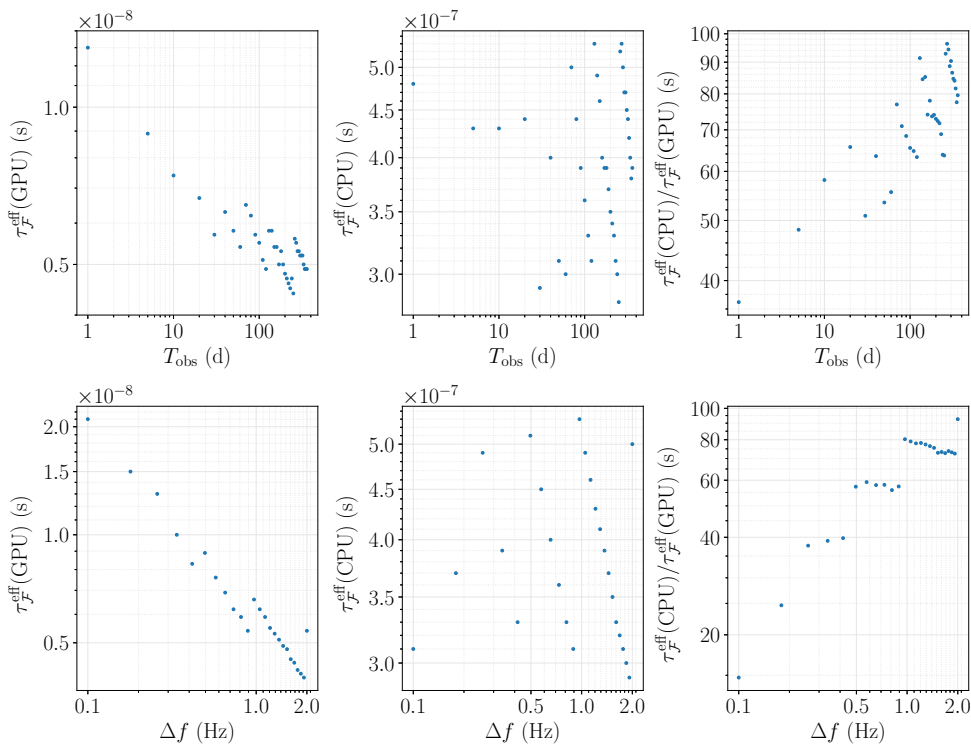}
    \caption{Comparative per-frequency bin runtime ($\tau_\mathcal{F}^\mathrm{eff}$ in s) of the CPU  (middle column) and GPU implementations (left column) of \texttt{Resamp}.  The right column displays the speed-up factor $\tau_\mathcal{F}^\mathrm{eff} (\mathrm{CPU}) / \tau_\mathcal{F}^\mathrm{eff} (\mathrm{GPU})$. \emph{Top row}: timing results as $T_\mathrm{obs}$ is varied, with $\Delta f = 0.606\,\mathrm{Hz}$. \emph{Bottom row}: timing results as $\Delta f$ is varied, with $T_\mathrm{obs} = 10\,\mathrm{d}$.}
    \label{fig:bench_res}
\end{figure}
The timing results are presented in figure \ref{fig:bench_res}.
The leftmost and middle panels give the absolute timing coefficients for the two implementations, while the rightmost panels show the speed-up factor $\tau_\mathcal{F}^\mathrm{eff} (\mathrm{CPU}) / \tau_\mathcal{F}^\mathrm{eff} (\mathrm{GPU})$

Figure \ref{fig:bench_res} indicates that between one and two orders of magnitude of acceleration is typical, with an overall trend towards higher $\tau_\mathcal{F}^\mathrm{eff} (\mathrm{CPU}) / \tau_\mathcal{F}^\mathrm{eff} (\mathrm{GPU})$ as the amount of data grows. 

The step jumps and subsequent declines in both the CPU and GPU timing results occur because the data are padded, such that the dimension of the data vector is a power of two, which greatly improves FFT performance.
The jumps occur, when the number of samples increases to the point, where the next power of two must be used, and there is the greatest mismatch between the original number of samples and the number of samples in the padded data.

Note that these timing results are averaged by searching over 100 sky locations for each choice of $T_\mathrm{obs}$ or $\Delta f$.
Seaching over spin-down parameters instead (e.g. $100$ values of $f_0^{(1)}$) leads to a ${\sim} 60\%$ reduction in the computation time for the CPU implementation, but only a ${\sim}20\%$ reduction in the computation time for the GPU implementation.
In both cases the reduction in computation time is due to re-use of the results of the sky position-dependent barycentering.
This barycentering consumes a larger fraction of the total computation time in the CPU case than in the GPU case.
As a result, the overall speed-up of the GPU implementation over the CPU implementation is reduced by a factor of ${\sim}2$ when searching only over spin-down parameters, holding $(\alpha, \delta)$ fixed.

We caution here that the values of $\tau^\mathrm{eff}_\mathcal{F}$ refer only to the time taken to \emph{calculate} $2\mathcal{F}$.
The computational cost of any processing of the calculated $2\mathcal{F}$ values (for example, sorting to obtain the top $1\%$ of candidates) is not included.
In particular, $\tau^\mathrm{eff}_\mathcal{F} \mathrm{(GPU)}$ does not include the time taken to transfer the calculated $2\mathcal{F}$ values from GPU memory back into CPU memory or to subsequently process those values using CPU-bound code.
Fortunately, the total amount of data to be transferred is not too great: the Nvidia P100 GPUs used in testing achieve a memory bandwidth of approximately $4.8\,\mathrm{GB}\,\mathrm{s}^{-1}$, and we have to transfer 4 bytes (one single-precision floating point number) per frequency bin, which costs an additional $8 \times 10^{-10}\,\mathrm{s}$ per frequency bin.
Compared to $\tau_\mathcal{F}^\mathrm{eff} \sim 5 \times 10^{-9}\,\mathrm{s}$, this is not negliglble, but it is a subdominant contribution to the overall computation time.
More costly is any processing which occurs in CPU-bound code: often this involves a serial loop over all frequency bins, which can be expensive.
For example, a simple loop which finds the maximum $2\mathcal{F}$ value in the frequency band adds $1.3 \times 10^{-9}\,\mathrm{s}$ per frequency bin.
This is also a subdominant contribution, but it is likely to increase if more complex processing is required, and so should be kept in mind.

\subsection{Accuracy}\label{subsec:acc}
\begin{figure}
    \centering
    \includegraphics[width=\columnwidth]{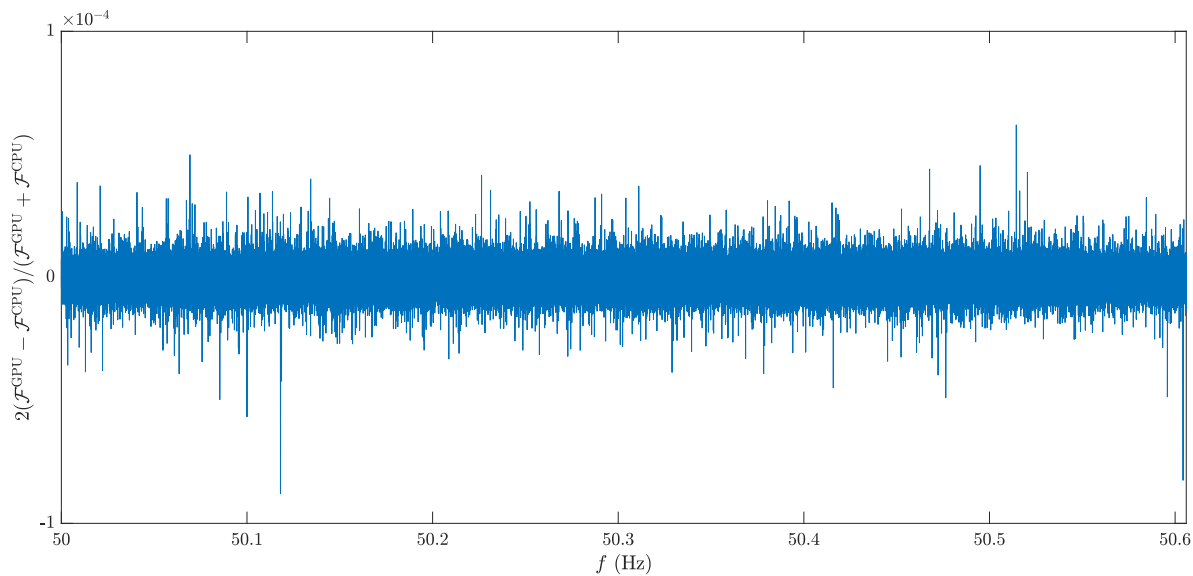}
    \caption{Fractional error between the $\mathcal{F}$-statistic computed with the GPU and CPU implementations of \texttt{Resamp} for a single noise realisation and as a function of search frequency $f$ (in Hz). Injected signal parameters: $h_0 = 6 \times 10^{-26},\, \alpha = 16^\mathrm{h}\, 19^\mathrm{m}\, 55.09^\mathrm{s}$, $\delta = -14^\circ\, 21'\, 35.1''$, $f_0 = 50.1\,\mathrm{Hz}$, $f_0^{(k)} = 0$ for $k \geq 1$.}
    \label{fig:manual_comp}
\end{figure}

As a first check of the accuracy of the GPU implementation, we manually compare the GPU and CPU output for a single noise realisation. 
The noise is white and Gaussian with one-sided amplitude spectral density $S_h(f)^{1/2} = 4 \times 10^{-24}\,\mathrm{Hz}^{-1/2}$. A continuous wave signal is injected at $50.1\,\mathrm{Hz}$ with wave strain $h_0 = 6 \times 10^{-26}$.
We use $10$ days of synthetic data.
The fractional discrepancy between the two methods is shown in Figure~\ref{fig:manual_comp}.
The fractional error in the $\mathcal{F}$-statistic does not exceed $1.5 \times 10^{-4}$ across the ${\sim} 10^6$ frequency bins in the range $50 \leq f_0 / (1\,\mathrm{Hz}) \leq 50.606$. The root-mean-square fractional error across the range is $3.7 \times 10^{-6}$.

To further check the accuracy of the GPU implementation, we employ an existing diagnostic in \texttt{LALSuite}, called \texttt{ComputeFstatTest}, which checks for consistency between different algorithms for computing $\mathcal{F}$, by varying three template parameters: binary orbital period $P$, right ascension $\alpha$, and first frequency derivative $f_0^{(1)}$.
The parameter ranges we check are recorded in table \ref{tbl:acc_params}.
\begin{table}
\centering
\begin{tabular}{lrl}\hline
    Parameter & Range & Bin spacing\\ \hline
    $P$ (hr) & $[0, 24]$ & 1.0   \\
    $\alpha$ (rad)  & $[0, 2\pi]$ &  $0.01 $ \\
    $f_0^{(1)}$ ($\mathrm{nHz}\,\mathrm{s}^{-1}$) & $[-2, -1]$ & $0.01$\\ \hline
\end{tabular}
\caption{Template parameter ranges used in the accuracy tests described in section \ref{subsec:acc}}
\label{tbl:acc_params}
\end{table}
Note that in these tests there is no noise injected, so that separate runs give the same results.
There is a continuous wave signal injected at $100\,\mathrm{Hz}$ with an unrealistically large strain $h_0 = 1$. 

\begin{figure}
    \centering
    \includegraphics[width=\columnwidth]{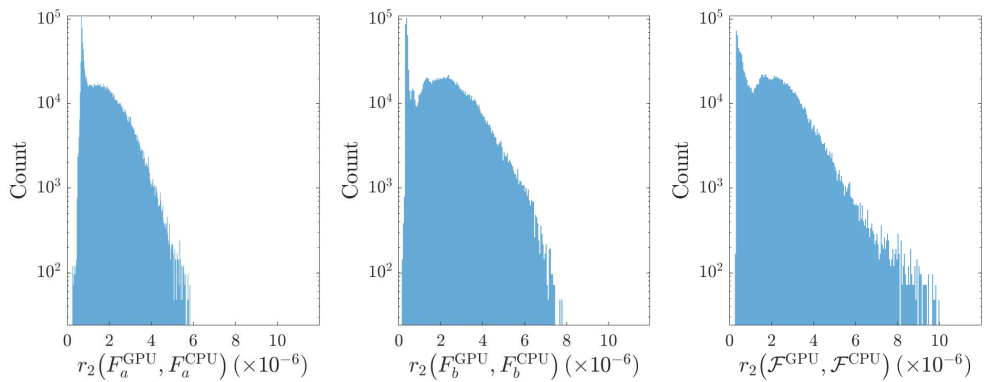}
    \caption{Histograms of the relative error in terms of the $L^2$ norm between GPU and CPU output for $F_a$ (\emph{left panel}), $F_b$ (\emph{middle panel}), and $\mathcal{F}$ \emph{(right panel}) for the \texttt{ComputeFstatTest} tests described in section \ref{subsec:acc}.
    Injected signal parameters: $h_0 = 1,\, \alpha = 0,\, \delta = -28^\circ\, 48',\, f_0 = 100\,\mathrm{Hz},\, f_0^{(1)} = -10^{-9}\,\mathrm{Hz}\,\mathrm{s}^{-1},\, f_0^{(k)} = 0$ for $k \geq 2$.}
    \label{fig:computefstattest}
\end{figure}

\texttt{ComputeFstatTest} enlists three comparison metrics to compare the output for $\mathcal{F}$, $F_a$, and $F_b$.
Viewing each of $\mathcal{F}, F_a, F_b$ as a real (in the case of $\mathcal{F}$) or complex (in the case of $F_a$ and $F_b$) vector with dimensionality equal to the number of frequency bins (1000 for these tests), we compute the relative error between the GPU and CPU results in terms of the $L^1$ and $L^2$ norms:
\begin{equation}
    r_{1,2}(x, y) = \frac{2\abs{x-y}_{1,2}}{\abs{x}_{1,2} + \abs{y}_{1,2}}.
\end{equation}
We also compute the angle between the two vectors as
\begin{equation}
    \theta = \arccos{\left[\frac{\Re(x^*\cdot y)}{\abs{x}_2\abs{y}_2}\right]},
\end{equation}
where $x^* \cdot y$ is the usual inner product.
For each set of template parameters we confirm that these metrics do not exceed the arbitrary default tolerances in \texttt{ComputeFstatTest} when comparing the existing CPU implementation of \texttt{Resamp} with the GPU implementation.
Histograms of the relative error between the GPU and CPU output with the $L^2$ norm for $F_a$, $F_b$, and $\mathcal{F}$ are shown in Figure \ref{fig:computefstattest}.
These show that the $L^2$ relative errors do not exceed ${\sim}10^{-5}$ in these tests.
For comparison, the default tolerances in \texttt{ComputeFstatTest} are of the order $10^{-2}$.
The results in Figure \ref{fig:computefstattest} are safely within the latter bounds.
We briefly note the apparent two-peaked structure in all three panels of Figure \ref{fig:computefstattest}.
We expect that this is a numerical artifact, but we are not able to conclusively identify a root cause for this behaviour.

\section{Pilot search}
\label{sec:pilot}
In this section we apply the GPU implementation of the $\mathcal{F}$-statistic to a pilot search. We focus on the acceleration attained by the GPU implementation rather than attempting to draw astrophysical conclusions; the design is a precursor to a future, more comprehensive study.
A full astrophysical search is outside the scope of this paper, but verifying the utility of a GPU-accelerated $\mathcal{F}$-statistic in a realistic search context is nonetheless a valuable exercise, and that is our aim here.
In Section \ref{subsec:pilot_targets} we describe the targets of the pilot search.
In Section \ref{subsec:pilot_data_params} we discuss the data to be used in the search, and the choice of search parameters.
In Section \ref{subsec:pilot_run_acc} we evaluate the runtime and accuracy of the GPU implementation under these realistic conditions, and find that the results are consistent with those in Sections \ref{subsec:runtime} and \ref{subsec:acc}.
Finally in Section \ref{subsec:pilot_cand} we discuss briefly the results of the search.

\subsection{Targets}
\label{subsec:pilot_targets}
The pilot search is a narrowband search for the four known isolated millisecond pulsars in the globular cluster Omega Centauri (NGC 5139), PSRs~J1326$-$4728A, C, D, and~E.
These pulsars were recently discovered, in 2018 and 2019, in a radio survey undertaken at the Parkes radio telescope following up on Australia Telescope Compact Array imaging of the core of Omega Centauri \cite{daiDiscoveryMillisecondPulsars2020}.
Of the four, only PSR~J1326$-$4728A has a complete timing solution with a well-constrained sky position $(\alpha, \delta)$ and spin-down rate $f^{(1)}_\star$, but the pulse frequency $f_\star$ of each pulsar is well-measured electromagnetically\footnote{The subscript $\star$ refers to values intrinsic to the rotation of the star, while the subscript $0$ refers to parameters related to the gravitational wave signal at the detector, as in previous sections.}. 
Being newly discovered, none of these pulsars have previously been the subject of a published directed gravitational wave search.

Globular cluster neutron stars have been targeted in several previous searches in LIGO data.
The nearby globular cluster NGC 6544 was the subject of a directed search for neutron stars not detected as pulsars \cite{abbottSearchContinuousGravitational2017}.
The search was carried out on $9.2$ days of data from the sixth LIGO science run, and searched coherently with the $\mathcal{F}$-statistic over a wide range of spin-down parameters ($\abs{f_0^{(1)}/f_0} \lesssim 10^{-10}\,\mathrm{s}^{-1}$).
Known pulsars in globular clusters have also been the subject of several targeted searches, e.g. \cite{abbottSearchesGravitationalWaves2010, aasiGravitationalWavesKnown2014, Abbott2019Known}.
In these cases the positions and spin-down parameters of the pulsars are well-measured electromagnetically, so the parameter space which needs to be covered is small and it is computationally feasible to use long ($> 100\,\mathrm{d}$) stretches of data in the search.
These targeted searches use coherent analysis methods including the $\mathcal{F}$-statistic as well as the Bayesian \cite{dupuisBayesianEstimationPulsar2005a} and $5n$-vector \cite{astoneMethodDetectionKnown2010a} methods.

The pilot search described here represents a middle ground between the above two approaches.
We are searching for four \emph{known} pulsars, but three of these pulsars have unconstrained spin-down parameters and large uncertainties on their sky position, forcing us to adopt an approach more akin to a directed search  (except that the spin frequency is well-constrained, obviating the need to search over a large number of frequency bands).
As noted above, this search is a validation exercise, whose goals are to assess the acceleration achieved by the GPU $\mathcal{F}$-statistic and its ease of use in a practical context. It is not a full astrophysical search. Nevertheless, the selected targets are well motivated astrophysically and could easily form part of a full  search in the future.

\subsection{Data and search parameters}
\label{subsec:pilot_data_params}
We perform the search using data from the second LIGO observing run (O2), which lasted from November 30, 2016 to August 25, 2017.
Strain data from the complete run are publicly available, but we do not search using the full dataset because the computational cost of a fully-coherent search over $(\alpha, \delta, f_0, f_0^{(1)})$ scales as $T_\mathrm{obs}^5$ \cite{bradySearchingPeriodicSources1998}.
We prefer to avoid including data which are of lower quality and do not contribute much to sensitivity.
As such, we perform a fully coherent search on the 100-day segment from January 3, 2017 (GPS time~$1\,167\,458\,304$) to April 13, 2017 (GPS time~$1\,176\,098\,304$).
During this time both the Hanford and Livingston detectors maintained high duty factors (fraction of time spent taking science-mode data) of $74\%$ and $66\%$ respectively, compared to $59\%$ and $57\%$ respectively across the entirety of O2.

For each pulsar we search $1\,\mathrm{Hz}$ sub-bands centred on $f_0 = f_\star$ and $2f_\star$, where $f_\star$ is the pulse frequency reported by Ref.~\cite{daiDiscoveryMillisecondPulsars2020}.
We use the default spacing between frequency templates of $1/(2T_\mathrm{obs}) = 5.7 \times 10^{-8}\,\mathrm{Hz}$.
\begin{table}
    \centering
    \begin{tabular}{cllll}\hline
        Pulsar & $f_0$ (Hz) & $f_0^{(1)}$ (Hz\,s$^{-1}$) & RAJ & DECJ \\\hline
        J1326-4728A & $[242.88, 243.88]$ & $[-1.8, -1.4] \times 10^{-15}$ & $13^\mathrm{h}26^\mathrm{m}39.7^\mathrm{s}$ & $-47^\circ30'11.64''$ \\
        & $[486.26, 487.26]$ & $[-3.6, -2.8] \times 10^{-15}$ & $13^\mathrm{h}26^\mathrm{m}39.7^\mathrm{s}$ & $-47^\circ30'11.64''$ \\
        J1326-4728C & $[145.1, 146.1]$ & $[-2.8, 0.33] \times 10^{-14}$ & $13^\mathrm{h}26^\mathrm{m}44^\mathrm{s} \pm 7'$ & $-47^\circ 29'40'' \pm 7'$ \\
        & $[290.7, 291.7]$ & $[-5.7, 0.66] \times 10^{-14}$ & $13^\mathrm{h}26^\mathrm{m}44^\mathrm{s} \pm 7'$ & $-47^\circ 29'40'' \pm 7'$ \\
        J1326-4728D & $[217.9, 218.9]$ & $[-3.0, 0.50] \times 10^{-14}$ & $13^\mathrm{h}26^\mathrm{m}44^\mathrm{s} \pm 7'$ & $-47^\circ 29'40'' \pm 7'$ \\
        & $[436.3, 437.3]$ & $[-6.0, 1.0] \times 10^{-14}$ & $13^\mathrm{h}26^\mathrm{m}44^\mathrm{s} \pm 7'$ & $-47^\circ 29'40'' \pm 7'$ \\
        J1326-4728E & $[237.16, 238.16]$ & $[-3.0, 0.54] \times 10^{-14}$ & $13^\mathrm{h}26^\mathrm{m}44^\mathrm{s} \pm 7'$ & $-47^\circ 29'40'' \pm 7'$ \\
        & $[474.8, 475.8]$ & $[-6.1, 1.1] \times 10^{-14}$ & $13^\mathrm{h}26^\mathrm{m}44^\mathrm{s} \pm 7'$ & $-47^\circ 29'40'' \pm 7'$\\ \hline
    \end{tabular}
    \caption{Search parameters for the pilot search described in Section \ref{sec:pilot}. Each pair of lines corresponds to $f_0 = f_\star$ and $2f_\star$.}
    \label{tbl:pilot_params}
\end{table}
For those pulsars which do not have an accurately measured position, we search over the half-power extent of the Parkes radio telescope pointing in which the pulsars were discovered, which measures approximately $14' \times 14'$ at an observing frequency of $1.4\,\mathrm{GHz}$ and is centred on $\mathrm{RAJ} = 13^\mathrm{h}26^\mathrm{m}44^\mathrm{s}$, $\mathrm{DECJ} = -47^{\circ}29'40''$ \cite{daiDiscoveryMillisecondPulsars2020}.

Three of the four pulsars also do not have a well-measured spin-down rate $f^{(1)}_\star$.
The observed value of $f^{(1)}_\star$ is due to both the intrinsic spin-down of the pulsar and line-of-sight acceleration of the pulsar in the gravitational potential of the globular cluster.
The intrinsic spin-down of millisecond pulsars $f^{(1)}_{*, \mathrm{int}}$ is no more than $-2.5 \times 10^{-14}\,\mathrm{Hz}\,\mathrm{s}^{-1}$ for all but the brightest millisecond pulsars, with radio luminosity $R_{1400} \gtrsim 50\,\mathrm{mJy}\,\mathrm{kpc}^2$ at $1.4\,\mathrm{GHz}$\cite{manchesterAustraliaTelescopeNational2005}.
PSRs J1326$-$4728C, D, and E are detected as relatively dim sources with $R_{1400} \lesssim 1.2\,\mathrm{mJy}\,\mathrm{kpc}^2$ \cite{daiDiscoveryMillisecondPulsars2020}.
We therefore regard a large intrinsic spin-down as unlikely for these pulsars, and do not consider values of $f^{(1)}_{\star, \mathrm{int}} < -2.5 \times 10^{-14}\,\mathrm{Hz}\,\mathrm{s}^{-1}$.
A rule of thumb for estimating the maximum apparent frequency derivative $f^{(1)}_{\star,\mathrm{a}}$ due to gravitational acceleration $a_\mathrm{max}$ is given by Ref. \cite{phinneyPulsarsProbesNewtonian1992}: \begin{equation} \left|\frac{f^{(1)}_{\star, \mathrm{a}}}{f_\star}\right| < \frac{a_\mathrm{max}}{c} \approx \frac{1}{c}\frac{\sigma(R_\bot)^2}{\sqrt{r_\mathrm{c}^2 + R_\bot^2}}, \end{equation} where $R_\bot$ is the projected distance of the pulsar to the cluster centre, $\sigma(R_\bot)$ is the line-of-sight velocity dispersion at $R_\bot$, and $r_\mathrm{c}$ is the core radius.
Given a central velocity dispersion $\sigma(0) = 22\,\mathrm{km}\,\mathrm{s}^{-1}$ and a core radius $r_\mathrm{c} = 4.6\,\mathrm{pc}$ \cite{meylanCentralVelocityDispersion1995} we set $R_\bot = 0$ and estimate $a_\mathrm{max}/c \approx 1.14 \times 10^{-17}\,\mathrm{s}^{-1}$.
This approximation is good to $\sim 50\%$ \cite{phinneyPulsarsProbesNewtonian1992}, so we take the upper bound on $\abs{f^{(1)}_{*,a}/f}$ to be $2 a_\mathrm{max}/c = 2.3 \times 10^{-17}\,\mathrm{s}^{-1}$ for safety.
The range of spin-down values to be considered for each pulsar without an existing measurement are then obtained by combining the limits on intrinsic spindown with the limits on spin-down due to gravitational acceleration: \begin{equation} -2.5 \times 10^{-14}\,\mathrm{Hz}\,\mathrm{s}^{-1} - \frac{2a_\mathrm{max}f_\star}{c} \leq f^{(1)}_\star \leq \frac{2a_\mathrm{max}f_\star}{c}. \end{equation}

Table \ref{tbl:pilot_params} lists the search parameters for each object.
The number of sky position and $f_0^{(1)}$ templates varies between frequency bands, with bands at higher frequency requiring more templates to cover the parameter space with a prescribed mismatch value $\mu$.
Here we take $\mu = 0.1$ corresponding to an upper bound on the reduction of the signal-to-noise ratio due to template mismatch of $10\%$.
For simplicity, the metric which controls the template spacing is an approximate metric introduced by \cite{bradySearchingPeriodicSources1998,Whit2006:ObsCnGrvWEmSpCS}, which is provided in \texttt{LALSuite}.
This metric accounts for modulation of the phase, but not the amplitude, of the signal due to detector motion, in contrast to the complete metric described by \cite{prixSearchContinuousGravitational2007}.
As the goal of this search is only to investigate the utility of computing $2\mathcal{F}$ using \texttt{CUDA}, we are content to make this simplifying approximation here.
The precise layout of the templates in the parameter space is also handled by the standard \texttt{LALSuite} routines.
In total, $5.51 \times 10^5$ sky position and spin-down templates are evaluated at $1.70 \times 10^7$ frequencies each, for a total of $9.36 \times 10^{12}$ computed $\mathcal{F}$-statistic values.
The number of sky position templates dominates the total number of templates: the search required $2.73 \times 10^5$ sky pointings, but only $11$ spin-down templates.
Note that the sky pointings and spin-down templates are not distributed evenly among the bands searched: bands at higher $f_0$ require more sky pointings and spin-down templates.
A single spin-down template is required for all bands except the $2f_\star$ bands of PSRs J1326$-$4728C, D, and E.

\subsection{Runtime and accuracy}
\label{subsec:pilot_run_acc}
The search takes $17.2$ GPU-hours on the same hardware as used in Section \ref{sec:perf}, corresponding to $\tau_\mathcal{F}^{\mathrm{eff}}(\mathrm{GPU}) = 6.6 \times 10^{-9}\,\mathrm{s}$.
For the purpose of comparison, we run the more expensive CPU implementation on a small subset of the parameter space covered in the search, and find $\tau_\mathcal{F}^\mathrm{eff}(\mathrm{CPU}) = 4.2 \times 10^{-7}\,\mathrm{s}$.
Based on this value of $\tau_\mathcal{F}^{\mathrm{eff}}(\mathrm{CPU})$, we extrapolate that the pilot search would consume roughly $1092$ CPU core-hours.
The overall speedup factor $\tau_\mathcal{F}^\mathrm{eff}(\mathrm{CPU})/\tau_\mathcal{F}^{\mathrm{eff}}(\mathrm{GPU})$ is 64.

These timing results probe roughly the same regime as the timing measurements shown in Figure \ref{fig:bench_res} and are consistent with those results.
Reading off Figure \ref{fig:bench_res}, we expect $\tau_\mathcal{F}^{\mathrm{eff}} (\mathrm{GPU}) \approx 5 \times 10^{-9}\,\mathrm{s}$ and $\tau_\mathcal{F}^{\mathrm{eff}} (\mathrm{CPU}) \approx 3.5 \times 10^{-7}\,\mathrm{s}$ for an overall speedup factor $\approx 70$.
We remind the reader that in practice the speedup of a complete search may not always be comparable to the speedup of the computation of $2\mathcal{F}$.
Once the computation of $2\mathcal{F}$ has been accelerated, other bottlenecks may emerge.
For example, if every computed $2\mathcal{F}$ value is to be retained on disk and/or processed further using CPU-bound code, then the search time may be dominated by the GPU to CPU memory transfer or disk I/O steps (see Section \ref{subsec:runtime}).
These new bottlenecks may themselves be ameliorated on a case-by-case basis (e.g. moving the CPU-bound post-processing step onto the GPU where practical, or employing non-blocking I/O to write the results to disk).

As in Section \ref{subsec:acc}, we verify the accuracy of the new implementation by comparison with the existing CPU-bound implementation of the resampling algorithm.
For PSRs J1326-4728C, D, and E, we choose a $1' \times 1'$ sky region at random from within the $14' \times 14'$ sky region which is searched, and output all $2\mathcal{F}$ values calculated with both the existing implementation and the \texttt{CUDA} implementation in the first $10^{-3}\,\mathrm{Hz}$ of the two frequency bands centred on $f_\star$ and $2f_\star$.
The sky region is chosen at random for each frequency band.
We verify that the fractional difference in $2\mathcal{F}$ values does not exceed $2.5 \times 10^{-4}$ and the root-mean-square fractional difference in each band does not exceed $8 \times 10^{-6}$.
These results, which we emphasise are performed on real rather than synthetic data, are similar to those found in the Gaussian noise tests of Section \ref{subsec:acc}.

\subsection{Candidates}
\label{subsec:pilot_cand}
While we do not seek to present a full astrophysical search here, for completeness we briefly discuss the results of the pilot search.
For the purposes of this exercise, we retain only those candidates with $2\mathcal{F} > 70$.
This retention step is implemented on the GPU side, so we do not incur the cost of transferring the full band back to the CPU and processing it serially there.
In pure Gaussian noise $2\mathcal{F}$ follows a $\chi^2$ distribution with 4 degrees of freedom \cite{Schutz1998}, and so the probability of at least one of the $N_\mathrm{t} = 9.36 \times 10^{12}$ templates exceeding the $2\mathcal{F}$ threshold is $1 - [\Pr(2\mathcal{F} < 70)]^{N_\mathrm{t}} \approx 0.2$.
This is therefore quite a conservative threshold, with a per-search false alarm probability of $20\%$ assuming Gaussian noise.
No 2$\mathcal{F}$ values exceeding $70$ are returned.
We do not convert these non-detections into astrophysical upper limits on the amplitude of CW signals from these targets, as this would require expensive Monte Carlo injections which are beyond the scope of this paper.

\section{Discussion}\label{sec:discussion}
We describe a new \texttt{CUDA} implementation for computing the $\mathcal{F}$-statistic via the resampling algorithm. The new implementation takes advantage of the power of GPUs to provide order-of-magnitude speed-ups over the existing CPU  code available in \texttt{LALSuite}.

The new code should not be viewed as a replacement for the existing methods. Rather, it works alongside them, allowing $\mathcal{F}$-statistic searches to take advantage of any GPUs at their disposal.
The new implementation is validated on synthetic data with Gaussian noise.
We find that the speedup with the new implementation scales with both the observing time $T_\mathrm{obs}$ and the size of the frequency band $\Delta f$, with typical speedup factors between 10 and 100.
We also verify that the \texttt{CUDA} implementation remains accurate: synthetic data tests indicate that the fractional difference between the new implementation and the existing CPU implementation does not exceed $1.5 \times 10^{-4}$.
Finally, we describe a pilot search using the new implementation on 100 days of real data from O2.
We target the four known isolated pulsars in the globular cluster Omega Centauri, which were discovered only recently \cite{daiDiscoveryMillisecondPulsars2020}.
These are all millisecond pulsars, with well-known pulse frequencies; but in three cases the spin-down frequency is unconstrained and the position is only known to an accuracy of $\pm7'$, necessitating a search over a significant number ($5.51 \times 10^5$) of sky position and spin-down combinations for these targets.
The search takes $17.2$ GPU-hours to complete compared to a projected $45.5$ CPU core-days, a speedup of $\approx 64$ times.
We verify that the discrepancies between the $2\mathcal{F}$ values computed by the \texttt{CUDA} implementation and the CPU implementation remain small ($\lesssim 2.5 \times 10^{-4}$ fractionally) in real data.
No detection candidates are found.
The pilot search demonstrates the utility of the new GPU     implementation in the context of a realistic search with a novel selection of targets, but we emphasise that it is not a substitute for a full astrophysical search, which is beyond the scope of this paper.

No effort has been made at this stage to perform optimisations in areas such as memory layout and transfer, removing conditionals within the \texttt{CUDA} kernels, or searching several templates simultaneously in cases where the GPU's capability is not saturated by the computation required for a single template.
The work reported here represents a first pass at the problem, with room for further optimisation in the future.

Beyond further optimisation, we briefly mention two other avenues for future work.
One is the comparison of this implementation with two other in-development implementations of the $\mathcal{F}$-statistic which employ \texttt{CUDA}\footnote{\url{https://github.com/mbejger/polgraw-allsky/tree/master/search/network/src-gpu}} and \texttt{OpenCL}\footnote{\url{https://github.com/MathiasMagnus/polgraw-allsky/tree/d36f9c238b76c2d683db393ee1cc34a2a3100bf1/search/network/src-opencl}} (M. Bejger, private communication).
Another is investigating whether this work can be used to speed up the resampling implementation \cite{Meadors2018} of the CrossCorrelation search for gravitational waves from sources in binary systems.

In~\cite{1HzSearch} we perform an all-sky search for continuous gravitational waves over a $1\,\mathrm{Hz}$ frequency band, pursuing a strategy of maximising sensitivity over a narrow parameter domain.
The search uses a semi-coherent search algorithm \cite{WettPrix2013:FPrmMtASrGrvPl, Wett2015:PrmMASmSrGrvPl, WettEtAl2018:ImpSmSCnGrvWUOCnTB}; data from O2 are partitioned into segments of duration $\sim 10\,\mathrm{d}$, each segment is searched using the \fstat{}, and results from all segments are then summed together.
GPUs are used to compute the \fstat{} for each segment, using the implementation presented here, as well as the sum over segments.
The search achieves a sensitivity of $h_0 = 1.01 \times 10^{-25}$ and is limited by available computational resources.
We estimate that, without the computational speed-up from the GPU \fstat{} implementation, but with the same limited computational resources, the sensitivity achieved after re-configuring the parameters of the search (number of segments, segment length, template placement, and frequency band) would have been reduced by $\sim 8\%$.
For example, the segment length would have been shortened to $\sim 5.5\,\mathrm{d}$ to reduce the computational cost of the \fstat{}.
The modest difference in sensitivity, with and without the speed-up from the GPU \fstat{} implementation, is ultimately due to the shallow scaling of sensitivity with computational cost for semi-coherent algorithms \cite{PrixShal2012:SCntGrvWOpStMFCmC}.
However, it is still significant scientifically, in situations where sound astrophysical arguments exist (e.g. indirect spin-down limits) that plausible sources are emitting near the detection limits of current searches \cite{2013riles}.

\ack
We thank David Keitel for helpful comments on the manuscript.
This research is supported by the Australian Research Council Centre of Excellence for Gravitational Wave Discovery (OzGrav) through project number CE170100004.
This work was performed on the OzSTAR national facility at Swinburne University of Technology. OzSTAR is funded by Swinburne University of Technology and the National Collaborative Research Infrastructure Strategy (NCRIS).

\section*{References}
\bibliography{gpufstat}
\end{document}